\documentclass [12pt]{article}
\usepackage{amsmath,amssymb,cite}

\usepackage{tabularx}
\usepackage[english]{babel}
\usepackage[dvips]{graphicx}
\usepackage{indentfirst}
\usepackage{epsfig}
\begin{document}

\title{A PROPOSAL FOR A NON-PERTURBATIVE REGULARIZATION OF ${\cal N}=2$ SUSY $4$D GAUGE THEORY}

\author{Badis Ydri\footnote{ydri@physik.hu-berlin.de. The Humboldt Universitat Zu Berlin preprint number is HU-EP-07/36.}\\
Institut fur Physik, Humboldt-universitat zu Berlin\\
Newtonstr.15, D-12489 Berlin-Germany.}

\maketitle

\begin{abstract}
In this letter we show that supersymmetry like geometry  can be approximated using finite dimensional matrix models and fuzzy manifolds. In particular we propose a non-perturbative regularization of ${\cal N}=2$ supersymmetric $U(n)$ gauge action in $4$D. In some planar large $N$ limits we recover exact SUSY together with the smooth geometry of ${\bf R}^4_{\theta}$.
\end{abstract}
%\tableofcontents

Noncommutative geometry \cite{CONNES} is the only known modification of field theory which preserves supersymmetry. In this note we will go one step further beyond the infinite dimensional matrix algebras of noncommutative Moyal-Weyl spacetimes and show that supersymmetry ( like geometry itself ) can be approximated using finite dimensional matrix models and fuzzy manifolds \cite{thesis,thesis1}. In particular we propose a non-perturbative regularization of ${\cal N}=2$ supersymmetric $U(n)$ gauge action in $4$ dimensions. In some planar large $N$ limits we will recover exact SUSY action together with the smooth geometry of spacetime ${\bf R}^4_{\theta}$.

Motivated by $1)$ the IKKT matrix models approach \cite{IKKT,IKKT1} to spacetime generation and $2)$ the noncommutative fuzzy geometry approach \cite{thesis,thesis1} to $i)$ quantum geometry  and to $ii)$ the non-perturbative quantum field theory we are led to the following considerations and a proposal for  a non-perturbative regularization of ${\cal N}=2$ SUSY in $4$ dimensions using finite dimensional $N\times N$ matrix algebras.

Let us now consider the following bosonic actions \cite{NCFQED} 
\begin{eqnarray}
S^{(X)}_B&=&N\bigg[-\frac{1}{4}Tr[X_a,X_b]^2+\frac{2i{\alpha}}{3}{\epsilon}_{abc}TrX_aX_bX_c\bigg].
\end{eqnarray}
\begin{eqnarray}
S^{(Y)}_B&=&N\bigg[-\frac{1}{4}Tr[Y_a,Y_b]^2+\frac{2i{\alpha}}{3}{\epsilon}_{abc}TrY_aY_bY_c\bigg].
\end{eqnarray}
\begin{eqnarray}
S^{(XY)}_B=-\frac{N}{2}Tr[X_a,Y_b]^2.
\end{eqnarray}
$X_a$ and $Y_a$ are $N\times N$ matrices where $N=(L+1)^2n$. We define  $g^2=1/(N^2{\alpha}^4)$. The two dimensional model given by the matrices $X_a$ alone is studied extensively in \cite{ref}. The corresponding action is some modification of $S^{(X)}_B$ which involves the addition of a potential term which is polynomial in $X_a^2$.

The minumum of the model is the solution of the conditions $F_{ab}=i[X_a,X_b]+\alpha {\epsilon}_{abc}X_c=0$, $G_{ab}=i[Y_a,Y_b]+\alpha {\epsilon}_{abc}Y_c=0$ and $H_{ab}=i[X_a,Y_b]=0$ . This solution is given explicitly by the matrices $X_a=\alpha L_a$ and $Y_a=\alpha K_a$ where $\{L_a\}$ and $\{K_a\}$ are two commuting sets of  generators of $SU(2)$ in the irreducible representation $\frac{L}{2}$ which satisfy $[L_a,L_b]=i{\epsilon}_{abc}L_c$, $[K_a,K_b]=i{\epsilon}_{abc}K_c$ and $[L_a,K_a]=0$. These matrices define the fuzzy $S^2\times S^2$ geometry \cite{NCFQED,madore,ref1}. Expanding around these matrices by writing $X_a=\alpha (L_a+A_a)$ and $Y_a=\alpha (K_a+B_a)$ and substituting back into the action $S^{(X)}_B+S^{(Y)}_B+S^{(XY)}_B$ we get a $U(n)$ gauge theory on fuzzy $S^2\times S^2$ with gauge coupling constant equal $g^2$. The $U(n)$ gauge transformations are implemented by $U(N)$ unitary matrices. The $A_a$ are the components of the gauge field in the directions of the first sphere while $B_a$ are the components in the directions of the second sphere. Two of these components are normal to the spheres and hence the true $4-$dimensional gauge field is also coupled to two scalar fields ( these two normal components ) which ( by construction ) transform in the adjoint representation of the group $U(N)$.

The crucial point to note here is the fact that without the Chern-Simons-like terms given by $iTrX_1X_2X_3$ and $iTrY_1Y_2Y_3$ in $S^{(X)}_B$ and $S^{(Y)}_B$ we will  get no finite dimensional useful geometry. By solving the equations of motion which are given in this case by $F_{ab}=i[X_a,X_b]=0$, $G_{ab}=i[Y_a,Y_b]=0$ and $H_{ab}=i[X_a,Y_b]=0$ we find that the minimum is given by diagonal matrices, in other words the geometry is trivial which is that of  a single point. We will also be able to get the Moyal-Weyl geometry  in the large $N$ limit in this model ( i.e the model without the Chern-Simons-like term ). This is because the Moyal-Weyl space ${\bf R}^2_{\theta}\times {\bf R}^2_{\theta}$  can not be   realized in terms of finite dimensional matrices. In  the presence of the Chern-Simons-like terms the Moyal-Weyl geometry can still be obtained in  large $N$ planar limits.
%where the conditions to be satisfied by the  minimum become $F_{ab}=i[X_a,X_b]=0$, $G_{ab}=i[Y_a,Y_b]=0$ and $H_{ab}=i[X_a,Y_b]=0$.

We introduce the following new variables $D_{\mu}=(D_1\equiv X_1,D_2\equiv X_2, D_3\equiv Y_1, D_4\equiv Y_2)$ and $X_3=\frac{\phi +{\phi}^+}{\sqrt{2}}$, $iY_3=\frac{{\phi}-{\phi}^+}{\sqrt{2}}$. Then we can write the bosonic action in the form
\begin{eqnarray}
S^{(X)}_B+S^{(Y)}_B+S^{(XY)}_B&=&\frac{N}{4}TrF_{\mu \nu}^2+NTr[D_{\mu},\phi]^+[D_{\mu},\phi]+NTr[\phi,{\phi}^+]^2\nonumber\\
&+&\sqrt{2}\alpha NTr\phi (F_{12}-iF_{34})+\sqrt{2}\alpha NTr{\phi}^+(F_{12}+iF_{34}).\label{4}
\end{eqnarray}
In above  the curvature tensor is defined now by $F_{\mu \nu}=i[D_{\mu},D_{\nu}]$. The Chern-Simons-like couplings ( which are strictly real here ) are given in the second line. The term $NTr[\phi,{\phi}^+]^2$ can be replaced by $-\frac{N}{2}TrD^2+NTr[\phi,{\phi}^+]D$ where we have now to do an extra integral over the hermitian $N\times N$ matrix $D$. The action becomes
\begin{eqnarray}
S^{(X)}_B+S^{(Y)}_B+S^{(XY)}_B&=&\frac{N}{4}TrF_{\mu \nu}^2+NTr[D_{\mu},\phi]^+[D_{\mu},\phi]-\frac{N}{2}TrD^2+NTr[{\phi},{\phi}^+]D\nonumber\\
&+&\sqrt{2}\alpha NTr\phi (F_{12}-iF_{34})+\sqrt{2}\alpha NTr{\phi}^+(F_{12}+iF_{34}).
\end{eqnarray}
We recognize this action ( modulo the second line ) to be the bosonic action of ${\cal N}=2$ SUSY $4$D $U(n)$ gauge theory where the ${\bf R}^4$ geometry is reduced to a point. Thus we know ( more or less ) what are the fermionic terms to be added to have full supersymmetry with the first line of this action. The Chern-Simons-like term will not be supersymmetrized. It is exactly these Chern-Simons-like terms which will provide in some approximate sense a non-trivial geometry which will resemble in some large $N$ limit the geometry of ${\bf R}^4$. It is in this large $N$ limit ( to be thought of as a continuum limit ) that we recover exact SUSY ( since this term becomes vanishingly small ) and also recover smooth ${\bf R}^4$. As it turns out we can implement full ${\cal N}=2$ SUSY with the first line even for finite matrix size $N$ by including the correct fermionic degrees of freedom of the ${\cal N}=2$ theory with the canonical "{\it gauge covariant }" supersymmetry transformations. We will follow the notation and convention of Weinberg with one exception which we will indicate at the end.

First we need to superymmetrize the action
\begin{eqnarray}
\frac{S^{(1)}_B}{N}&=&\frac{1}{4}TrF_{\mu \nu}^2-\frac{1}{2}TrD^2.
\end{eqnarray}
We add the gaugino action
\begin{eqnarray}
\frac{S^{(1)}_F}{N}&=&\frac{1}{2} Tr\bar{\lambda}{\gamma}^{\mu}[D_{\mu},\lambda].
\end{eqnarray}
The gaugino field $\lambda$ is a $4N\times N$ matrix with Grassmann  matrix elements. This is a Majorana field. $D_{\mu}$, $D$ and ${\lambda}$ are members of the same  supersymmetric multiplet with transformation properties
\begin{eqnarray}
&&{\delta}D_{\mu}=i\bar{\epsilon}{\gamma}_{\mu}\lambda \nonumber\\
&&{\delta}D=i\bar{\epsilon}{\gamma}_5{\gamma}_{\mu}[D_{\mu},\lambda]\nonumber\\
&&{\delta}{\lambda}=\frac{1}{4}F_{\mu \nu}[{\gamma}_{\mu},{\gamma}_{\nu}]\epsilon -i{\gamma}_5D\epsilon~,~{\delta}\bar{\lambda}{\equiv}({\delta}{\lambda})^+{\beta}=-\frac{1}{4}\bar{\epsilon}[{\gamma}_{\mu},{\gamma}_{\nu}]F_{\mu \nu} -i\bar{\epsilon}{\gamma}_5D.
\end{eqnarray}
${\delta}D_{\mu}$ and ${\delta}D$ are $N\times N$ hermitian matrices while $({\delta}{\lambda})_{\alpha}$ and $({\delta}\bar{\lambda})_{\alpha}$ are $N\times N$ matrices with Grassmann matrix elements.
The variation of the bosonic action is
\begin{eqnarray}
\frac{{\delta}S^{(1)}_B}{N}=iTr{\delta}D_{\mu}[D_{\nu},F_{\mu \nu}]-TrD{\delta}D.
\end{eqnarray}
The variation of the fermionic action is
\begin{eqnarray}
\frac{S^{(1)}_F}{N}&=&\frac{1}{2} Tr{\delta}\bar{\lambda}{\gamma}^{\mu}[D_{\mu},\lambda]+\frac{1}{2} Tr\bar{\lambda}{\gamma}^{\mu}[D_{\mu},\delta\lambda]\nonumber\\
&=&\frac{1}{4}Tr(\bar{\epsilon}[{\gamma}_{\mu},{\gamma}_{\nu}]{\gamma}^{\rho}\lambda ) [D_{\rho},F_{\mu \nu}]+iTrD[D_{\rho},\bar{\epsilon}{\gamma}_5{\gamma}^{\rho}\lambda ]\nonumber\\
%&=&\frac{1}{4}Tr(\bar{\epsilon}[{\gamma}_{\mu},{\gamma}_{\nu}]{\gamma}^{\rho}\lambda ) [D_{\rho},F_{\mu \nu}]+iTrD[D_{\rho},\bar{\epsilon}{\gamma}_5{\gamma}^{\rho}\lambda ]\nonumber\\
&=&\frac{1}{4}Tr(\bar{\epsilon}[{\gamma}_{\mu},{\gamma}_{\nu}]{\gamma}^{\rho}\lambda ) [D_{\rho},F_{\mu \nu}]+TrD{\delta}D.
\end{eqnarray}
In the first line above we have used the fact that $Tr\bar{\lambda}{\gamma}^{\mu}[\delta D_{\mu},\lambda ]=0$ which is due to the identity that given any Majorana field $\lambda$ we have $\bar{\lambda}{\gamma}^{\mu}\lambda =0$. We have the identity $[{\gamma}^{\mu},{\gamma}^{\nu}]{\gamma}^{\rho}=-2{\eta}^{\mu \rho}{\gamma}^{\nu}+2{\eta}^{\nu \rho}-2i{\epsilon}^{\mu \nu \rho \sigma}{\gamma}_{\sigma}{\gamma}_5$. The last term leads to the  Jacobi identity ${\epsilon}^{\mu \nu \rho \sigma}[D_{\rho},[D_{\mu},D_{\nu}]]=0$ whereas the other two terms lead to the result
\begin{eqnarray}
\frac{{\delta}S^{(1)}_F}{N}=-iTr{\delta}D_{\mu}[D_{\nu},F_{\mu \nu}]+TrD{\delta}D.
\end{eqnarray}
Hence $S^{(1)}_B+S^{(1)}_F$ is supersymmetric as expected. We need now to add the other members of the ${\cal N}=2$ supermultiplet. The fields $D_{\mu}$, $D$ and $\lambda$ form an ${\cal N}=1$ gauge supermultiplet. The ${\cal N}=2$ supermultiplet will also contain an ${\cal N}=1$ chiral supermultiplet with components $\phi$ ( the above scalar field ), ${\psi}$ ( another Majorana field ) and $F$ ( the chiral multiplet's auxilary field ). Following Weinberg we will impose an extra R-symmetry relating the two Majorana spinors $\lambda$ and $\psi$ via the transformation $\psi {\longrightarrow}\lambda$, $\lambda {\longrightarrow}-\psi$ and hence the two $N=1$ supermultiplets will naturally form an $N=2$ supermultiplet.

Thus the goal now is to supersymmetrize the following bosonic action
\begin{eqnarray}
\frac{S^{(2)}_B}{N}&=&Tr[D_{\mu},\phi]^+[D_{\mu},\phi]+Tr[{\phi},{\phi}^+]D.
\end{eqnarray}
We will set the auxilary field $F$ to zero from the start. This is in anyway the value at which the ${\cal N}=2$ action is stationary. 
The variation of the above bosonic action under some SUSY transformations of fields is given by %( with $D=2[{\phi},{\phi}^+]$ )

\begin{eqnarray}
\frac{{\delta}S^{(2)}_B}{N}&=&Tr{\delta}D_{\mu}\bigg([[D_{\mu},{\phi}^+],{\phi}]+[[D_{\mu},\phi],{\phi}^+]\bigg)+Tr[\phi,{\phi}^+]{\delta}D\nonumber\\
&+&Tr{\delta}{\phi}\bigg[[D_{\mu},[D_{\mu},{\phi}^+]]-Tr[D,{\phi}^+]\bigg]+Tr{\delta}{\phi}^+\bigg[[D_{\mu},[D_{\mu},{\phi}]]+Tr[D,{\phi}]\bigg].
\end{eqnarray}
%The supersymmetric variations ${\delta}D_{\mu}$ of the $N\times N$ hermitian matrices $D_{\mu}$ are matrices which will be allowed for consistency to be given by general $N\times N$ matrices. As it turns out we will require anti-hermitian variations $({\delta}D_{\mu})^+=-{\delta}D_{\mu}$.

Motivated by the canonical ${\cal N}=2$ supersymmetry in $4$ dimensions we try the following fermionic terms
\begin{eqnarray}
\frac{S^{(2)}_F}{N}&=&a Tr\bar{\psi}{\gamma}^{\mu}[D_{\mu},\psi]+b\bigg(Tr\bar{\psi}_L[\phi, {\lambda} ] -Tr\bar{\lambda}[{\phi}^+,{\psi}_L]\bigg).
\end{eqnarray}
This action is real because ${\psi}_L$ and $\lambda$ are Grassmann. Indeed because $\psi$ is Garssmann and because $\beta ({\gamma}^{\mu})^{+}\beta =-{\gamma}^{\mu}$ we have $(\bar{\psi}_L{\gamma}^{\mu}[D_{\mu},{\psi}_L])^{+}=-[D_{\mu},\bar{\psi}_L{\gamma}^{\mu}]{\psi}_L$ and hence $(Tr\bar{\psi}{\gamma}^{\mu}[D_{\mu},\psi])^*=Tr\bar{\psi}{\gamma}^{\mu}[D_{\mu},\psi]$. Similar argument holds for the other two terms where we will find that we need the above relative minus sign to get a real action. As we have already said 
$\lambda$ is a Majorana fermion while $\psi$ is defined now as the Majorana fermion whose left-handed component is given by ${\psi}_L$. Thus the kinetic term should be rewritten
\begin{eqnarray}
Tr\bar{\psi}{\gamma}^{\mu}[D_{\mu},\psi ]=Tr\bar{{\psi}_L}{\gamma}^{\mu}[D_{\mu},{\psi}_L]-Tr[D_{\mu},{\bar{\psi}_L}]{\gamma}^{\mu}{\psi}_L.
\end{eqnarray}
We assume the following extra SUSY transformations

\begin{eqnarray}
&&{\delta}\phi =i\sqrt{2}\bar{\epsilon}{\psi}_L~,~{\delta}{\phi}^+=i\sqrt{2}\bar{{\psi}_L}\epsilon  \nonumber\\
&&{\delta}{\psi}_L =i\sqrt{2}[D_{\mu},\phi]{\gamma}^{\mu}{\epsilon}_R~,~{\delta}\bar{{\psi}_L}{\equiv}{\delta}{\psi}_L^+\beta  =-i\sqrt{2}[D_{\mu},{\phi}^+]\bar{{\epsilon}_R}{\gamma}^{\mu}.
\end{eqnarray}
 $\delta \phi$ , $\delta {\phi}^+$ are $N\times N$ complex matrices while $({\delta}{\psi}_L)_{\alpha}$ $({\delta}\bar{{\psi}_L})_{\alpha}$ are $N\times N$ matrices with Grassmann entries. We have the identities
\begin{eqnarray}
&&Tr\bar{\psi}{\gamma}^{\mu}[{\delta}D_{\mu},\psi ]=0\nonumber\\
&&Tr\bar{{\psi}_L}[{\delta}\phi,\lambda ]-Tr\bar{\lambda}[{\delta}{\phi}^+,{\psi}_L]=0.
\end{eqnarray}
Thus the variation of the fermionic action under SUSY transformations is
\begin{eqnarray}
\frac{{\delta}S_F^{(2)}}{N}&=&aTr\delta{\bar{\psi}_L}{\gamma}^{\mu}[D_{\mu},{\psi}_L]+aTr\bar{{\psi}_L}{\gamma}^{\mu}[D_{\mu},\delta{\psi}_L]-aTr[D_{\mu},\delta{\bar{\psi}_L}]{\gamma}^{\mu}{\psi}_L-aTr[D_{\mu},\bar{{\psi}_L}]{\gamma}^{\mu}\delta{\psi}_L\nonumber\\
&+&bTr{\delta}\bar{{\psi}_L}[\phi,\lambda ]-bTr{\delta}\bar{\lambda}[{\phi}^+,{\psi}_L]+bTr\bar{{\psi}_L}[\phi,{\delta}\lambda]-bTr\bar{\lambda}[{\phi}^+,{\delta}{\psi}_L].
\end{eqnarray}
\begin{eqnarray}
{\rm 1st~line}&=&2\sqrt{2}iaTr(\bar{\epsilon}{\gamma}^{\nu}{\gamma}^{\mu}{\psi}_L) [D_{\mu},[D_{\nu},{\phi}^+]]+2\sqrt{2}iaTr(\bar{{\psi}_L}{\gamma}^{\mu}{\gamma}^{\nu}\epsilon )[D_{\mu},[D_{\nu},\phi]]\nonumber\\
&=&-\frac{a}{\sqrt{2}}Tr(\bar{\epsilon}[{\gamma}^{\mu},{\gamma}^{\nu}]{\psi}_L)[F_{\mu \nu},{\phi}^+]-2aTr{\delta}\phi [D_{\mu},[D_{\mu},{\phi}^+]]\nonumber\\
&+&\frac{a}{\sqrt{2}}Tr(\bar{{\psi}_L}[{\gamma}^{\mu},{\gamma}^{\nu}]{\epsilon})[F_{\mu \nu},{\phi}]-2aTr{\delta}{\phi}^+ [D_{\mu},[D_{\mu},{\phi}]].
\end{eqnarray}
Also ( using the fact that $\bar{{\epsilon}_R}{\gamma}^{\mu}\lambda=\bar{\epsilon}{\gamma}^{\mu}{\lambda}_R$ and $\bar{\lambda}{\gamma}^{\mu}{\epsilon}_R=-\bar{\epsilon}{\gamma}^{\mu}{\lambda}_L$ )
\begin{eqnarray}
bTr{\delta}\bar{{\psi}_L}[\phi,\lambda ]-bTr\bar{\lambda}[{\phi}^+,{\delta}{\psi}_L]&=&-\frac{b}{\sqrt{2}}Tr{\delta}D_{\mu}\bigg[[[D_{\mu},{\phi}^+],\phi]+[[D_{\mu},\phi],{\phi}^+]\bigg]-\frac{b}{\sqrt{2}}Tr{\delta}D[{\phi},{\phi}^+].\nonumber\\
%-ib\sqrt{2}Tr(\bar{\epsilon}{\gamma}^{\mu}{\lambda}_L)[D_{\mu},[{\phi},{\phi}^+]]\nonumber\\
%&=&-\frac{b}{\sqrt{2}}Tr{\delta}D_{\mu}[D_{\mu},[\phi,{\phi}^+]]-\frac{b}{\sqrt{2}}Tr{\delta}D[\phi,{\phi}^+].
\end{eqnarray}
\begin{eqnarray}
-bTr{\delta}\bar{\lambda}[{\phi}^+,{\psi}_L]+bTr\bar{{\psi}_L}[\phi,{\delta}\lambda]&=&\frac{b}{4}Tr(\bar{\epsilon}[{\gamma}_{\mu},{\gamma}_{\nu}]{\psi}_L)[F_{\mu \nu},{\phi}^+]-\frac{b}{4}Tr(\bar{{\psi}_L}[{\gamma}_{\mu},{\gamma}_{\nu}]{{\epsilon}})[F_{\mu \nu},{\phi}]\nonumber\\
&+&\frac{b}{\sqrt{2}}Tr{\delta}{\phi}[D,{\phi}^+]-\frac{b}{\sqrt{2}}Tr{\delta}{\phi}^+[D,\phi].
\end{eqnarray}
We verify quite easily that with the values $b=\sqrt{2}$, $a=1/2$ we will have ${\delta}S^{(2)}_F=-{\delta}S^{(2)}_B$.

The full action is

\begin{eqnarray}
\frac{1}{N}\bigg(S^{(X)}_B+S^{(Y)}_B+S^{(XY)}_B\bigg)_{\rm SUSY}&=&\frac{1}{4}TrF_{\mu \nu}^2+Tr[D_{\mu},\phi]^+[D_{\mu},\phi]-\frac{1}{2}TrD^2+Tr[{\phi},{\phi}^+]D\nonumber\\
&+&\frac{1}{2} Tr\bar{\lambda}{\gamma}^{\mu}[D_{\mu},\lambda]+\frac{1}{2} Tr\bar{\psi}{\gamma}^{\mu}[D_{\mu},\psi]\nonumber\\
&+&\sqrt{2}\bigg(Tr\bar{\psi}_L[\phi, {\lambda} ] -Tr\bar{\lambda}[{\phi}^+,{\psi}_L]\bigg)\nonumber\\
&+&\sqrt{2}\alpha Tr\phi (F_{12}-iF_{34})+\sqrt{2}\alpha Tr{\phi}^+(F_{12}+iF_{34}).
\end{eqnarray}

The first three lines constitute the full ${\cal N}=2$ SUSY $U(n)$ 
gauge theory on a single point. The last line ( although it breakes explicitly SUSY ) is added so to be able to have a well defined finite dimensional geometry on which the theory lives. This term will also allow us to have a rigorous continuum limit. In some appropriate "planar" limit this term will go to zero and hence we recover exact SUSY as well as a smooth geometry. This is another way of getting SUSY on Moyal-Weyl spaces. Let us explain this point a little further. We write the above action in the following way

\begin{eqnarray}
\frac{1}{N}\bigg(S^{(X)}_B+S^{(Y)}_B+S^{(XY)}_B\bigg)_{\rm SUSY}&=&\frac{1}{4}Tr\tilde{F}_{\mu \nu}^2+Tr[D_{\mu},\phi]^+[D_{\mu},\phi]-\frac{1}{2}TrD^2+Tr[{\phi},{\phi}^+]D\nonumber\\
&+&\frac{1}{2} Tr\bar{\lambda}{\gamma}^{\mu}[D_{\mu},\lambda]+\frac{1}{2} Tr\bar{\psi}{\gamma}^{\mu}[D_{\mu},\psi]\nonumber\\
&+&\sqrt{2}\bigg(Tr\bar{\psi}_L[\phi, {\lambda} ] -Tr\bar{\lambda}[{\phi}^+,{\psi}_L]\bigg)\nonumber\\
&-&2{\alpha}^2Tr(X_3^2+Y_3^2).\label{line}
%&+&\sqrt{2}\alpha Tr\phi (F_{12}-iF_{34})+\sqrt{2}\alpha Tr{\phi}^+(F_{12}+iF_{34}).
\end{eqnarray}
In above $\tilde{F}_{12}=F_{12}+2\alpha X_3=-\tilde{F}_{21}$, $\tilde{F}_{34}=F_{34}+2\alpha Y_3=-\tilde{F}_{43}$, $\tilde{F}_{13}=F_{13}=-\tilde{F}_{31}, \tilde{F}_{14}=F_{14}=-\tilde{F}_{41},\tilde{F}_{23}=F_{23}=-\tilde{F}_{32},\tilde{F}_{24}=F_{24}=-\tilde{F}_{42}$.
To study the noncommutative planar limit we should consider adding  to this action the following potential term 
\begin{eqnarray}
V[X_3,Y_3]&=&-Nm^2{\alpha}^2 TrX_3^2+\frac{2m^2}{N} Tr(X_3^2)^2 -Nm^2{\alpha}^2 TrY_3^2+\frac{2m^2}{N} Tr(Y_3^2)^2.
\end{eqnarray}
This potential is gauge invariant but not rotationally invariant. In above $m=N^p$ with some positive integer power $p$ so in the large $N$ limit we can see that this potential implements the constraints $X_3=\frac{N\alpha}{2}$  and $Y_3=\frac{N\alpha}{2}$ which means that on each sphere we are restricted to the north pole  in a covariant way. In this large $N$ limit if we also take ${\alpha}{\longrightarrow}0$ such that $N{\alpha}^2=1/{\theta}^2$ is kept fixed then we will obtain the noncommutative Moyal-Weyl plane with exact SUSY, viz
\begin{eqnarray}
\frac{1}{N}\bigg(S^{(X)}_B+S^{(Y)}_B+S^{(XY)}_B\bigg)_{\rm SUSY}&=&\frac{1}{4}Tr\tilde{F}_{\mu \nu}^2+Tr[D_{\mu},\tilde{\phi}]^+[D_{\mu},\tilde{\phi}]-\frac{1}{2}TrD^2+Tr[\tilde{\phi},\tilde{\phi}^+]D\nonumber\\
&+&\frac{1}{2} Tr\bar{\lambda}{\gamma}^{\mu}[D_{\mu},\lambda]+\frac{1}{2} Tr\bar{\psi}{\gamma}^{\mu}[D_{\mu},\psi]\nonumber\\
&+&\sqrt{2}\bigg(Tr\bar{\psi}_L[\tilde{\phi}, {\lambda} ] -Tr\bar{\lambda}[\tilde{{\phi}}^+,{\psi}_L]\bigg).
\end{eqnarray}
In above we have used the fact that the last line in (\ref{line}) leads to a constant  term in this planar limit. We have  also the definitions $\tilde{\phi}=\phi- \frac{N\alpha}{2}\frac{1+i}{\sqrt{2}}$, $\tilde{F}_{12}=F_{12}+\frac{1}{{\theta}^2}$,  $\tilde{F}_{34}=F_{34}+\frac{1}{{\theta}^2}$. The trace $Tr$ is now infinite dimensional. Under SUSY transfomations we will have the variations ${\delta}\tilde{\phi}={\delta}\phi$ ,
${\delta}\tilde{F}_{\mu \nu}={\delta}F_{\mu \nu}$ and hence this action is still ${\cal N}=2$ supersymmetric.

\paragraph{Remark :} The factor of $i$ in ${\delta}D_{\mu}$, ${\delta}\phi$ and ${\delta}{\phi}^+$ is due to our basic idenity which is given any pair of Majorana spinors $s_1$ and $s_2$ ( which are here $4N\times N$ matrices ) we  have
\begin{eqnarray}
(\bar{s}_1Ms_2)^+=-\bar{s}_1Ms_2~,~M=1,{\gamma}_{\mu},[{\gamma}_{\mu},{\gamma}_{\nu}]
\end{eqnarray}
\begin{eqnarray}
(\bar{s}_1Ms_2)^+=+\bar{s}_1Ms_2~,~M={\gamma}_5,{\gamma}_{\mu}{\gamma}_5.
\end{eqnarray}
The signs in these two equations are opposite to the signs of equations $(26.{\rm A}.20)$ and $(26.{\rm A}.21)$ of Weinberg. In our case when we take the hermitian adjoint we include a minus sign ( in accordance with the property used in showing the reality of our action ) then when we reverse the interchange of $s_1$ and $s_2$ we get a second minus sign which cancels the first one.

\paragraph{Acknowledgements}
The work of Badis Ydri is supported by a Marie Curie Fellowship
from The Commission of the European Communities ( The Research
Directorate-General ) under contract number MIF1-CT-2006-021797.

\bibliographystyle{unsrt}

\end{document}